\documentstyle[11pt,paspconf,epsf]{article}

\begin{document}

\title{Measurements of the UV Upturn in Local and Intermediate-Redshift 
Ellipticals}
\author{Thomas M. Brown}
\affil{NOAO - NASA/GSFC}

\def\lesssim{\mathrel{\hbox{\rlap{\hbox{\lower4pt\hbox{$\sim$}}}\hbox{$<$}}}}
\def\gtrsim{\mathrel{\hbox{\rlap{\hbox{\lower4pt\hbox{$\sim$}}}\hbox{$>$}}}}

\begin{abstract}

The rest-frame ultraviolet contains the most sensitive indicators of
age for elliptical galaxies.  While the near-UV flux from young
ellipticals isolates the main sequence turnoff, the far-UV flux in old
ellipticals is dominated by hot horizontal branch (HB) stars.  This
evolved population was first revealed by early UV observations showing
a sharp flux increase shortward of rest-frame 2500~\AA, subsequently
dubbed the ``UV upturn.'' The phenomenon has since been characterized
in many local ellipticals, and measurements at intermediate redshifts
are now underway. Once ellipticals reach ages of 5--10~Gyr, stellar
and galactic evolution theories predict that the UV-to-optical flux
ratio can increase by orders of magnitude over timescales of a few
Gyr, making the UV upturn the most rapidly evolving feature of these
galaxies.  It is thus expected to fade dramatically with increasing
redshift.

I review the imaging and spectroscopic evidence for the nature of the
UV upturn in nearby ellipticals, and then present observations that
measure the UV upturn at an epoch significantly earlier than our
own. Far-UV data from the Hopkins Ultraviolet Telescope demonstrate
that the spectra of nearby ellipticals are dominated by hot HB stars.
Faint Object Camera UV imaging of M32 and the M31 bulge detected the
UV-bright phases of post-HB stars, but did not reach the HB
itself. Recent Space Telescope Imaging Spectrograph observations were
the first to image the hot HB and post-HB stars in the center of the
nearest elliptical galaxy, M32; these observations also show a
striking lack of UV-bright post-AGB stars. Faint Object Camera
observations of Abell~370, a rich galaxy cluster at $z=0.375$, show
that giant ellipticals at a lookback time of 4~Gyr can exhibit strong
UV luminosity, with no evidence of evolution in the UV upturn between
this epoch and our own, thus implying a high redshift of formation
($z_f \ge 4$).

\end{abstract}


\section{Introduction}

The UV upturn is a sharp rise in the spectra of elliptical galaxies
and spiral bulges shortward of rest-frame 2700~\AA.  Also known as the
``UV rising branch'' or the ``UV excess,'' it was discovered with the
OAO-2 satellite (Code 1969).  Prior to this discovery, we did not
expect such a UV-bright component to these old, passively-evolving
populations. Many candidates were suggested to explain the upturn,
including young massive stars, hot white dwarfs, hot horizontal branch
(HB) and post-HB stars, and non-thermal activity (see Greggio \&
Renzini 1990 for a complete review).  As the measurements of local
ellipticals were expanded with IUE, another surprise was the large
variation in the strength of the UV upturn from galaxy to galaxy, even
though the spectra of ellipticals appear very similar at longer
wavelengths. Characterized by the $1550-V$ color, the UV upturn
becomes stronger and bluer as the metallicity (optical Mg$_2$ index)
of the galaxy increases, while other colors become redder (Burstein et
al. 1988)

Today, it is widely believed that HB stars and their progeny are
responsible for the far-UV emission in elliptical galaxies. There are
three classes of post-HB evolution, each evolving from a different
range of effective temperature on the zero-age HB.  On the red end of
the HB, stars will evolve up the asymptotic giant branch (AGB),
undergo thermal pulses, evolve as bright post-AGB stars to hotter
temperatures, possibly form planetary nebulae, and eventually descend
the white dwarf (WD) cooling curve.  At hotter temperatures (and lower
envelope masses) on the HB, stars will follow post-early AGB
evolution: they evolve up the AGB, but leave the AGB before the
thermal pulsing phase, continue to high temperatures at high
luminosity, and descend the WD cooling curve. For the blue extreme of
the HB, stars with very little envelope mass will follow
AGB-Manqu$\acute{\rm e}$ evolution, evolving directly to hotter
temperatures and brighter luminosities without ever ascending the AGB,
and finally descending the WD cooling curve.  These three classes of
post-HB behavior have very different lifetimes, in the sense that the
post-AGB stars are bright in the UV for a brief period ($\sim 10^3 -
10^4$ yr), the post-early AGB stars are UV-bright for a longer period
($\sim 10^4-10^5$ yr), and the AGB-Manqu$\acute{\rm e}$ stars are
UV-bright for very long periods ($\sim 10^6-10^7$ yr).  The HB phase
itself lasts $\sim 10^8$~yr, and so the presence of hot HB stars in a
population, combined with the long-lived luminous post-HB phases,
gives rise to the strong UV upturn seen in the most massive,
metal-rich ellipticals (see Dorman, O'Connell, \& Rood 1995 and
references therein). A more significant fraction of the far-UV flux
from galaxies with a weak UV upturn can theoretically come from
post-AGB stars, and in the weakest UV upturn galaxies (e.g., M32), the
spectra alone do not require the presence of a hot HB.

The zero age HB (ZAHB) is not only a sequence in effective
temperature: it is also a sequence in mass (see Dorman, Rood, \&
O'Connell 1993).  For $\rm T_{eff} \gtrsim 6000$~K, a small change in
envelope mass ($\lesssim 0.1$~$M_\odot$) corresponds to a large change
in T$_{\rm eff}$ ($\Delta \rm T_{eff} \gtrsim 10,000$~K).  Because the
main-sequence turnoff (MSTO) mass decreases as age increases, the ZAHB
will become bluer as a population ages, assuming all other parameters
(e.g., mass loss on the red giant branch, metallicity, helium
abundance, etc.) remain fixed.  Note that this does not necessarily
imply that age is the ``second parameter'' of HB morphology (e.g., see
Sweigart's work in these proceedings).  The implied first parameter of
the HB morphology debate is metallicity; the HB becomes bluer at lower
metallicity, assuming all other parameters (age, mass loss, etc.)
remain fixed.  This is due to two reasons.  First, the MSTO mass at a
given age is lower at lower metallicity, because a metal-poor star is
more luminous (and thus shorter-lived) than a metal-rich star of the
same mass.  Second, as the metallicity decreases, the envelope opacity
decreases, the star will have a higher T$_{\rm eff}$, and the HB
becomes bluer.  In short, HB morphology tends to become bluer at lower
metallicity and higher ages, but other parameters also play a role
(rotation, He abundance, deep mixing, etc.), leading to the second
parameter debate. Our understanding of the UV upturn is clearly linked
to this debate, given the presence of hot HB stars in metal-rich
elliptical galaxies.

Because HB morphology is sensitive to age, Greggio \& Renzini (1990)
noted that the UV upturn should also be very sensitive to age;
Bressan, Chiosi, \& Fagotto (1994) demonstrated that the UV upturn
might even be useful as a diagnostic for determining the ages of old
ellipticals.  Indeed, using the infall model of chemical evolution,
Tantalo et al.\ (1996) demonstrated that the $1550-V$ color in an old
elliptical galaxy spectrum (age $\sim 6$~Gyr) could change by 4 mag
over a timescale of a few Gyr, while other well-studied colors ($V-K$,
$V-J$, $B-V$, $V-R$, $U-B$, etc.) only change by $\sim 0.1$ mag on the
same timescale.  Unfortunately, the many parameters that govern HB
morphology remain poorly constrained, and thus this age diagnostic has
yet to be calibrated.

\section{HUT Observations}

In 1990 and 1995, the Hopkins Ultraviolet Telescope (HUT) observed six
quiescent local elliptical galaxies as part of the Astro-1 and Astro-2
Space Shuttle missions (see Davidsen et al.\ 1992 for a description of
the instrument and its capabilities).  These galaxies -- NGC1399, M60,
M89, M49, NGC3115, and NGC3379 -- span a wide range of UV upturn
strength ($2.05 \le 1550-V \le 3.86$), but their far-UV
(900--1800~\AA) spectral energy distributions (SEDs) are all similar
to those of stars with 20,000~K $\leq$~T$_{\rm eff} \leq$ 23,000~K
(Brown, Ferguson, \& Davidsen 1995).  The data are inconsistent with
ongoing star formation following a normal IMF: there is a lack of
C{\small IV} absorption, and the flux decreases shortward of 1100~\AA\
(Ferguson et al.\ 1991). The HUT spectra are not dominated by a
population of post-AGB stars, else they would appear much hotter.
Furthermore, the fuel-consumption theorem (Greggio \& Renzini 1990)
relates the stellar evolutionary flux (SEF), also known as the stellar
death rate, to the bolometric luminosity within the HUT aperture; this
constraint means that the relatively short-lived post-AGB stars cannot
produce the magnitude of flux observed by HUT. However, the HUT
spectra are consistent with the integrated light of hot HB stars and
their descendants.  The data demonstrate that a minority population
($\lesssim 20$\% of the SEF) of hot HB stars and their descendants can
produce the far-UV light, with a small contribution from the remaining
stars that evolve along post-AGB tracks (Brown et al.\ 1997).

\section{Ultraviolet Imaging of M32}

M32 is usually considered a ``compact elliptical'' instead of a dwarf
elliptical (see Da Costa 1997 and references therein), and as such it
is the closest example of a true elliptical galaxy.  Given its
proximity, we can resolve the hot population in the core of the galaxy
with the HST.  Several programs have used the Faint Object Camera
(FOC) toward this end, prior to and after refurbishment of the HST
(see Brown et al.\ 1998b and references therein).  Brown et al.\
(1998b) resolved 183 stars in the center of M32; the colors and
luminosities of these stars placed them in the AGB-Manqu$\acute{\rm
e}$ and post-early-AGB evolutionary phases.  Although these UV
observations were not deep enough to resolve the hot HB, they did
imply that the Space Telescope Imaging Spectrograph (STIS) could
achieve this goal in a reasonable allocation of observing time.

In October 1998, the STIS observed a $25 \arcsec \times 25 \arcsec$
field, centered $7.7\arcsec$ from the M32 core, using the near-UV
detector (NUV-MAMA with F25QTZ filter), for 23 ksec. The relatively
large throughput of the bandpass, combined with the lack of red leak,
allowed the detection of 8000 hot HB and post-HB stars in the center
of M32.  Photometric reduction was done via PSF-fitting with the
DAOPHOT package (Stetson 1987); the stars were placed on the STMAG
photometric standard, where a flat $F_\lambda$ spectrum of 1 erg
s$^{-1}$ cm$^{-2}$ \AA$^{-1}$ corresponds to $-21.1$ mag. In the STMAG
magnitude system, the stars in our photometric catalog span a range of
21--28 mag; in the deepest parts of the image, the catalog is
reasonably complete ($>$ 25\%) to 27 mag.  The hot HB spans a range of
25--27 mag at T$_{\rm eff} >$ 8500~K, and thus these images are
sufficiently deep for photometry on the hot HB.  We have calculated a
set of solar-metallicity HB and post-HB evolutionary tracks for
comparison with the STIS data.

Although the analysis of the data is in progress, the southwest half
of the image, a section that ranges 8--20$\arcsec$ from the center of
the galaxy, is sufficiently deep and uncrowded to allow a preliminary
investigation of the evolved population.  In that region, the
uncorrected STIS luminosity function ranges from $\sim 10$ stars per
0.5~mag bin at 22--24 mag, and increases to hundreds of stars per bin
at magnitudes fainter than 25.  Including a completeness correction,
there are over a thousand stars per bin near 27 mag.

Post-AGB stars alone cannot produce such a luminosity function, if
they evolve in the manner described by the H-burning and He-burning
tracks found in the literature (e.g., Sch$\ddot{\rm o}$nberner 1987;
Vassiliadis \& Wood 1994).  In this portion of the STIS image, the
bolometric luminosity can be determined from the extant optical data,
a bolometric correction of $-0.875$ mag (Worthey 1994), an extinction
correction assuming $E(B-V)=0.11$ (Ferguson \& Davidsen 1993), and a
distance modulus of 24.43 mag. The fuel consumption theorem (Greggio
\& Renzini 1990) relates the bolometric luminosity to the stellar
evolutionary flux: SEF$= 2.2\times 10^{-11}$ stars yr$^{-1}
L_{\odot}^{-1} = 8.62\times 10^{-4}$ stars yr$^{-1}$.  Even with all
of these stars evolving through a given post-AGB channel, the post-AGB
tracks with mass larger than 0.6 M$_{\odot}$ evolve rapidly, and would
only produce a tiny fraction of the stars in the STIS image.  The
slowest tracks, such as the 0.546 M$_{\odot}$ track of Sch$\ddot{\rm
o}$nberner (1987), actually present two problems.  First, they produce
dozens of stars per 0.5~mag bin at 21 $\le$ STMAG $\le$ 24 mag, which
are many more than observed by STIS.  Second, this track also produces
dozens of stars per bin at 24 $\le$ STMAG $\le$ 28 mag, where STIS
observed an order of magnitude more, even without correcting for
completeness.  Because the post-AGB stars in this old population are
presumably evolving along low-mass tracks, the lack of bright stars in
the STIS image may be an indication that the post-AGB transition from
the AGB to the later and hotter phases (T$_{\rm eff} > 60,000$~K)
occurs on much more rapid timescales than those expected from the
canonical tracks, or is dust-enshrouded (see, e.g., K$\ddot{\rm
a}$ufl, Renzini, \& Stanghellini, 1993).

In contrast, the presence of hot HB stars and their progeny easily
reproduces the large number of faint stars seen by STIS.  If $\sim
5$\% of the SEF passes through the hot HB, the resulting luminosity
function is in close agreement with the STIS data, over the entire
range of observed magnitudes.  The resulting UV spectrum from such a
population agrees with the weak UV flux measured by IUE (Burstein et
al.\ 1988). Given the small fraction of the population entering the
hot HB, we must conclude that the vast majority of the population
passes through the red HB phase and the subsequent post-AGB evolution.
So, although a small population of hot HB stars can explain the stars
present in the STIS image, we still must explain the lack of bright
stars. The post-AGB stars must either be evolving along more massive
(and thus more rapidly evolving) tracks, the transition time from the
AGB to the hotter post-AGB phases must be more rapid than that
expected from the canonical low-mass tracks, or this transition must
be enshrouded in circumstellar dust.  Whichever the case, the
contribution of post-AGB stars to the STIS luminosity function would
be small.  Note that very young ages ($\lesssim 1$~Gyr) are required
to produce main sequence stars brighter than the HB, and so the STIS
image is dominated by an old population.

The STIS image convincingly demonstrates that the weak UV upturn in
M32 comes from a small population of hot HB stars and their
descendants. Prior to these data, the weak UV upturn in M32 could have
been explained by low-mass post-AGB stars alone, without violating
fuel consumption constraints on the SEF.  Far-UV imaging of the same
field with STIS would be useful to provide color information for these
stars.  Colors would further constrain the HB morphology, and provide
clues to the post-AGB evolution, because post-AGB stars presumably
contribute a small but significant fraction of stars to the STIS
luminosity function.

\section{Measurements of the UV Upturn in Abell 370}

The UV upturn is thought to evolve rapidly as a population ages.  Some
groups (e.g., Yi et al.\ 1999) have predicted a rapid decline in UV
upturn strength at increasing redshift; others (e.g. Tantalo et al.\
1996) predict that the UV upturn should show a strong increase at
intermediate ages ($\sim 6$ Gyr), followed by a period where it is
approximately constant.  In this latter scenario, the amount of fading
expected at higher redshifts depends upon the formation redshift of
the elliptical galaxy in question.  Several previous attempts to
measure the UV upturn at intermediate redshifts, using UV imaging and
spectroscopy with the HST, have been inconclusive (e.g., Windhorst et
al.\ 1994; Renzini 1996; Buson et al.\ 1998).

The FOC can observe with two long-pass filters that are appropriate
for observations of the UV upturn at intermediate
redshift. Specifically, the F130LP and F370LP filters provide a
discriminator between flux longward and shortward of 3700~\AA,
corresponding to a rest-frame wavelength of 2500--2700~\AA\ in the
redshift range 1.48--1.37. This rest-frame wavelength represents the
spectral division between the hot evolved populations that produce the
far-UV flux and the cooler RGB, AGB, and MSTO populations responsible
for the $V$-band flux.  We have obtained FOC measurements of four
ellipticals in the galaxy cluster Abell~370, at $z=0.375$, to measure
the UV upturn at a lookback time of $\sim 4$~Gyr (Brown et al.\
1998a).

Abell 370 is a well-studied galaxy cluster. The four ellipticals in
our analysis (BO\# 10, 24A, 24B, \& 34; Butcher, Oemler, \& Wells
1983) have all been observed spectroscopically from the ground
(Soucail et al.\ 1988); these spectra confirm their cluster
membership, are consistent with passively-evolving old populations,
and show no evidence of star formation (e.g., from [O{\small II}]).
Their $U-V$ colors are also consistent with passive evolution
(MacLaren, Ellis, \& Couch 1988).  Morphological classification is
confirmed by WFPC (Couch et al.\ 1994) and WFPC2 (HST GO program 6003)
imaging.

We performed aperture photometry upon these four cluster ellipticals,
using two aperture sizes: 65 pixels ($0.91\arcsec$) and 12 pixels
($0.17\arcsec$).  The large aperture includes all of the light
detected by the FOC; the smaller aperture gives a measurement of the
core light, and may be more appropriate for comparison to measurements
of the UV upturn in local ellipticals, which were done with IUE.
Local ellipticals usually show a weaker UV upturn at increasing radius
(Ohl et al.\ 1998).  We show our measured flux ratios in Table 1.
Exposure times varied from 2800--8500 sec, and so our counting
statistics are quite good.

\begin{table}
\caption{FOC Photometry and Model Predictions}
\begin{tabular}{|l|c|c|c|c|}
\tableline
          &   F130LP                 &      F370LP              &
            & rest-frame \\
Galaxy & (cts s$^{-1}) $ & (cts s$^{-1}) $ & F130LP/F370LP & 
$1550-V$ (mag) \\ 
\tableline \tableline
\multicolumn{5}{|c|}{Photometry: large aperture} \\
\tableline
BOW \#10 & 9.03 & 6.82 & 1.32 & 3.0 \\
BOW \#24a & 4.55 & 3.69 & 1.23 & 3.4 \\
BOW \#24b & 4.22 & 3.14 & 1.34 & 2.9 \\
BOW \#34 & 7.13 & 5.75 & 1.24 & 3.4 \\
\tableline
\multicolumn{5}{|c|}{Photometry: nuclear aperture}\\
\tableline
BOW \#10 & 1.78 & 1.21 & 1.47 & 2.2 \\
BOW \#24a & 0.841 & 0.650 & 1.29 & 3.1 \\
BOW \#24b & 1.11 & 0.840 & 1.32 & 2.9 \\
BOW \#34 & 1.48 & 1.16 & 1.28 & 3.2 \\
\tableline
\multicolumn{5}{|c|}{Redshifted non-evolving templates: nuclear aperture}\\
\tableline
NGC 1399 & 1.55 & 1.02 & 1.52 & 2.05 \\
M 60 & 1.26 & 0.856 & 1.47 & 2.24 \\
M 49 & 1.22 & 0.995 & 1.23 & 3.42 \\
\tableline
\end{tabular}
\end{table}

To determine the relationship between rest-frame $1550-V$ color and
the F130LP/F370LP flux ratio at this redshift, we first compared our
data to the spectra of local galaxies.  After redshifting the spectra
of three local ellipticals (M49, M60, and NGC1399) to $z=0.375$, we
used the IRAF/synphot/calcphot package to predict the F130LP/F370LP
flux ratio that would be observed for these local galaxies, if they
were members of Abell~370.  The predictions are shown in Table 1.  As
the table demonstrates, in a large aperture, the Abell 370 ellipticals
all show a UV upturn at least as strong as that seen in M49.  In a
nuclear aperture, three of these four ellipticals show even stronger
UV upturn colors, with one as strong as that in M60.

Comparison to the redshifted spectra of local ellipticals assumes a
non-evolving template for the interpretation.  However, the main
sequence turnoff becomes bluer and brighter at decreasing age, and so
we also compare our measurements with the predictions for an evolving
elliptical galaxy.  Tantalo et al.\ (1996) computed the SEDs of
elliptical galaxies as a function of age, assuming a range of galaxy
masses, with chemical evolution following the infall model. In their
models, the rise of the UV upturn occurs at $\sim 6$~Gyr, but note
that the actual age of this occurrence may vary by several Gyr,
depending upon the assumptions made regarding mass loss, $\Delta Y /
\Delta Z$, and details regarding the chemical evolution. The observed
F130LP/F370LP flux ratio translates to a rest-frame FUV-LP/NUV-LP flux
ratio at $z=0.375$, and so we calculated this rest-frame flux ratio as
a function of age for the two most massive models in the Tantalo et
al.\ (1996) set: $3\times 10^{12}$~$M_{\odot}$ and $1\times
10^{12}$~$M_{\odot}$.  The range of FUV-LP/NUV-LP flux ratio spanned
by these models is shown in Figure 1, plotted as a function of
lookback time and redshift, for three different formation redshifts
(2, 4, and 8).

\begin{figure}[h] \caption{The evolution of the UV upturn as a function of
redshift, as characterized by the rest-frame FUV-LP/NUV-LP flux ratio
in the SEDs of Tantalo et al.\ (1996) for giant ellipticals.  This
flux ratio corresponds to the observed FOC F130LP/F370LP ratio for
observations at $z=0.375$, the redshift of Abell 370.  Our
large-aperture measurements for Abell 370 (circles) show no evidence
of evolution in the UV upturn between a lookback time of 4~Gyr and
today, implying a high formation redshift ($z > 4$). }
\epsfxsize=4.75in \epsfbox{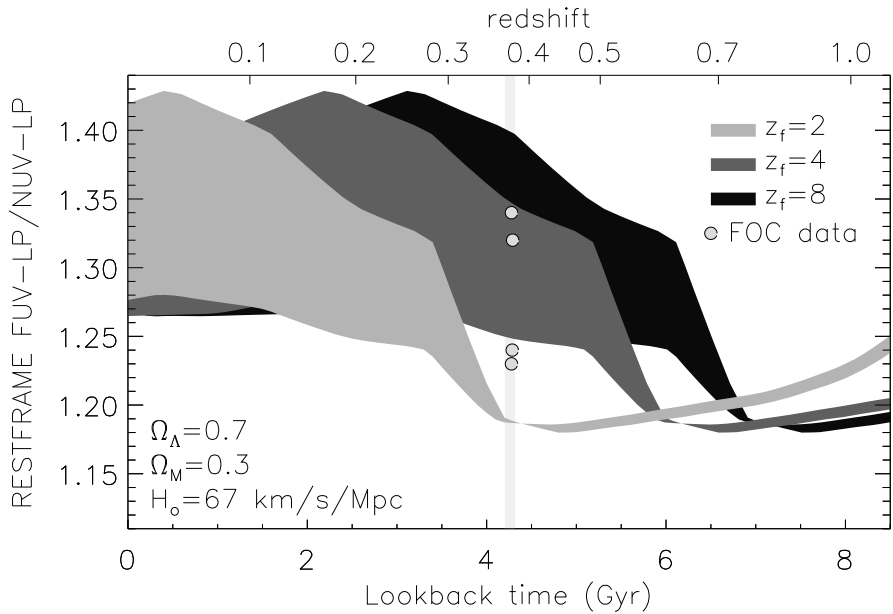} \end{figure}

Although our understanding of the UV upturn and HB morphology is not
sufficient to allow its use as an accurate age indicator, it is
interesting to consider the implications of our FOC measurements,
under the assumption that the Tantalo et al.\ (1996) models are
correct.  In the given cosmology, the ellipticals in Abell 370 would
have formed at $z> 4$ in order to demonstrate such strong UV upturn
colors.  For $\Omega_\Lambda = 0$ and $\Omega_M=0.1$, the implied
formation redshift is not significantly changed; for $\Omega_\Lambda =
0$ and $\Omega_M=0.3$, the implied formation redshift is higher. If
the Abell 370 ellipticals and the local population of ellipticals both
formed at a common redshift $z \ge 4$, the lack of evolution seen in
the UV upturn can be understood as evidence that both epochs are on
the ``flat'' portion of the UV upturn evolutionary curves, after the
UV upturn has leveled off.

Note that a small amount of star formation is sufficient to produce a
significant UV upturn in these galaxies.  Although these ellipticals
appear to be passively evolving, less than 0.1 $M_{\odot}$ yr$^{-1}$
of star formation would be enough to produce the UV upturn we observe
(see Madau, Pozzetti, \& Dickinson 1998); this level of star formation
is slightly below the constraints available from the extant data.

Our measurements provide a first step in mapping the evolution of the
UV upturn with lookback time. Further observations are clearly needed
to rule out star formation as the source of the UV emission in these
high redshift galaxies, and to trace the evolution to both higher and
lower redshifts.  HST Observations of a cluster at $z = 0.55$ are
planned in the near future (GTO program \#8020).  Unless the redshift
of formation is very high, these galaxies ought to be very faint in
the far-UV.  Once the UV upturn is measured in a statistically
significant number of galaxies over a range of redshifts, the
evolution can be determined independent of star formation
contamination. It would be unlikely that small levels of star
formation could be synchronized within clusters and change as a
function of redshift to mimic the UV upturn evolution.

\section{Summary}

The UV upturn provides a sensitive tracer of age in old populations,
and can potentially constrain the evolutionary history of ellipticals
with a diagnostic that is independent of those used at longer
wavelengths.  Spectra and photometry of local ellipticals confirm that
hot HB stars are responsible for the UV upturn, and so the calibration
of the UV upturn as an age diagnostic will depend upon a better
understanding of the mechanisms driving HB morphology (e.g., RGB mass
loss). Measurements of the UV upturn at a series of redshifts should
provide insight into the chemical evolution of galaxies and the
production of hot HB stars in metal-rich populations, with the
eventual goal of constraining the formation history of ellipticals.

\acknowledgements

The author is grateful for the collaborating efforts of C. W. Bowers
(NASA/GSFC), A. F. Davidsen (JHU), J.-M. Deharveng (MarsLab),
H. C. Ferguson (STScI), R. I. Jedrzejewski (STScI), R. A. Kimble
(NASA/GSFC), S. A. Stanford (IGPP/LLNL), and A. V. Sweigart
(NASA/GSFC).  Support for this work was provided by NASA through grant
GO-6667 and GO-5435 from the Space Telescope Science Institute, NAS
5-27000 to the Johns Hopkins University, and NAS 5-6499D to the
Goddard Space Flight Center.

\end{document}